\documentclass[a4paper, 11pt]{PoS}
\usepackage{amsmath}
\usepackage{slashed}

\title{Twist Three Generalized Parton Distributions for Orbital Angular Momentum}

\ShortTitle{Twist Three Generalized Parton Distributions for Orbital Angular Momentum}

\author{\speaker{Abha Rajan}\thanks{This work is supported by U.S. Department of Energy, Office of Science, Office of Nuclear Physics contracts DE-AC05-06OR23177, DE-FG02-01ER41200}\\
        University of Virginia\\
        E-mail: \email{ar5xc@virginia.edu}}
\author{Simonetta Liuti\\
        University of Virginia and INFN, Laboratori Nazionali di Frascati, Italy\\
        E-mail: \email{sl4y@virginia.edu}}

\abstract{We study the orbital angular momentum contribution to the spin structure of the proton. It is well known that the quark and gluon spin contributions do not add up to the proton spin. We motivate the connection between the Generalized Transverse Momentum Distribution (GTMD) $F_{14}$, and orbital angular momentum by exploring the underlying quark proton helicity amplitude structure. The twist three Generalized Parton Distribution (GPD) $\tilde{E}_{2T}$, was shown  to connect to OAM. We study these functions using a diquark model calculation. The GTMD $F_{14}$ is unique in that it can describe both Jaffe-Manohar and Ji OAM depending on choice of gauge link, {\it i.e.} whether final state interactions are included or not. We perform a calculation of $F_{14}$ in both scenarios. }

\FullConference{QCD Evolution 2015\\

                 May 26-30, 2015\\

                 Jefferson Lab (JLAB), Newport News Virginia, USA}

\begin{document}


\section{Introduction}

\noindent Orbital angular momentum is generated by the movement of quarks and and gluons inside the proton and is one of the candidates for accounting for the missing proton spin. Two famous decompositions of proton spin a la Jaffe Manohar \cite{JaffeManohar}
\begin{equation}
  \frac{1}{2} = \frac{1}{2}\Delta\Sigma + \mathcal{L}_q + \Delta G + \mathcal{L}_g  
\end{equation}
 and Ji \cite{Ji}
\begin{equation}
  \frac{1}{2} = J_q + J_g
\end{equation}
differ primarily in their definition of OAM $\mathcal{L}_{JM} \rightarrow i \vec{r}\times\vec{\partial}$ and $\mathcal{L}_{Ji} \rightarrow i \vec{r}\times\vec{\mathcal{D}}$. Defining OAM using the covariant derivative allows the Ji decomposition to be gauge invariant. In this description, the Generalized Parton Distributions (GPDs) $H$ and $E$ give access to the total partonic angular momentum J. 
\begin{equation}
  J_q = \frac{1}{2}\int_{-1}^1dx x(H_q(x,0,0) + E_q(x,0,0))
\end{equation}    

\noindent To access Ji's OAM, the spin contribution $\frac{1}{2}\Delta\Sigma$ needs to be subtracted. 
  
\begin{align}
  L_{Ji}  &= \frac{1}{2}\int_{-1}^1dx x(H_q(x,0,0) + E_q(x,0,0)) - \frac{1}{2}\int_{-1}^1dx \tilde{H}_q(x,0,0)\\
         &= J_q - \frac{1}{2}\Delta\Sigma 
\end{align}    

\noindent GPDs do not include transverse momentum and are defined in a collinear framework. Hence, $\mathcal{L}_{Ji}$ includes a straight gauge link. However, to describe $\mathcal{L}_{JM}$ we need to include final state interactions i.e. use a staple gauge link \cite{burkardt_staple}.

\section{Twist Two and Twist Three Helicity Amplitudes}
\noindent The quark quark correlator that describes these GPDs has the form,
\begin{equation}
W^{\gamma^+}_{\Lambda \Lambda'} = \int \frac{dz_-}{2 \pi} e^{i xP^+z^-} \langle p', \Lambda' \mid \bar{\psi}(-z/2)\gamma^+\psi(z/2)\mid p,\Lambda\rangle_{z^+=z_T=0} 
\end{equation}
In what follows, $p=P+\Delta/2$, $p'=P-\Delta/2$, $P= (p+p')/2$, $k=\bar{k}+\Delta/2$, $k'=\bar{k}-\Delta/2$, $\bar{k}= (k +k')/2$, and the skewness parameter, $\xi=\Delta^+/P^+ =0$, hence $t=\Delta^2$ ($t \equiv - {\Delta}_T^2$ for $\xi=0$).
\noindent This correlator can equivalently be described using helicity amplitudes $A_{\Lambda,\lambda,\Lambda',\lambda'}$ where $\Lambda$ ($\Lambda'$) and $\lambda$ ($\lambda'$) give proton and quark initial (final) helicities respectively. The proton helicities depend on the initial and final states chosen. The quark helicities, on the other hand, are dependent on the projection operator used. Hence, the general form of the correlator is given by,
\begin{equation}
W^{\Gamma}_{\Lambda \Lambda'} = \int \frac{dz_-}{2 \pi} e^{i xP^+z^-} \langle p', \Lambda' \mid \bar{\psi}(-z/2)\Gamma\psi(z/2)\mid p,\Lambda\rangle_{z^+=z_T=0} 
\end{equation}

\noindent At leading twist, $\Gamma$ : $\gamma^+$, $\gamma^+\gamma^5$, $i\sigma^{i+}$ give rise to the vector, axial vector (chiral even/ quark helicity non flip) and the tensor GPDs (chiral odd / quark helicity flip). This can also be understood by connecting to the helicity projection operator $1 \pm \gamma^5$ \cite{Die_rev}, 
\begin{equation}
A_{\Lambda,\pm,\Lambda',\pm} =\int \frac{dz_-}{2 \pi} e^{i xP^+z^-} \langle p', \Lambda' \mid \bar{\psi}(-z/2)\gamma^+(1 \pm \gamma^5)\psi(z/2)\mid p,\Lambda\rangle_{z^+=z_T=0} 
\end{equation}
\noindent The vector operator $\gamma^+$ ensures no spin flip. This gives us the definition of the leading twist non-flip helicity amplitudes.  $W^{\gamma^+}_{\Lambda,\Lambda'}$ and $W^{\gamma^+\gamma^5}_{\Lambda,\Lambda'}$ can be expressed in terms of helicity amplitudes as :

\begin{align}
W^{\gamma^+}_{\Lambda,\Lambda'} &= \frac{1}{2}(A_{\Lambda,+,\Lambda',+} + A_{\Lambda,-,\Lambda',-} ) \\
W^{\gamma^+\gamma^5}_{\Lambda,\Lambda'} &= \frac{1}{2}(A_{\Lambda,+,\Lambda',+} - A_{\Lambda,-,\Lambda',-} )
\end{align}

\noindent In the case of $\gamma^+$, as we are summing over the quark helicities, the quarks are unpolarized, while in the case of $\gamma^+\gamma^5$, the quarks are polarized. We also clearly see here how the Parton Distribution Function (PDF) $g_1(x)$ that parametrizes $W^{\gamma^+\gamma^5}$ in the forward case gives the quark spin contribution to the proton spin by looking at the underlying helicity amplitude structure:

\begin{equation}
\Delta \Sigma \rightarrow A_{++,++} - A_{+-,+-} - A_{-+,-+} + A_{--,--}
\end{equation}
\noindent i.e. longitudinally polarized quark in a longitudinally polarized proton. 

\noindent Now, if we think about the orbital angular momentum contribution, we need the quarks to be unpolarized \cite{LorcePas,us} (If they are polarized, the quark spin would contribute as well.) Hence, we would \textit{add} the quarks with opposite helicities. The amplitudes would look like 
\begin{equation}
\mathcal{L} \rightarrow A_{++,++} + A_{+-,+-} - A_{-+,-+} - A_{--,--} \label{oameq}
\end{equation}
At leading twist no PDF, GPD or TMD survives this combination of the amps. However if we include the partonic transverse momentum in the picture,
\begin{equation}
\mathcal{W}^{\gamma^+} = \int \frac{dz_-}{2 \pi} e^{i xP^+z^- -i k_T.z_T} \langle p', \Lambda' \mid \bar{\psi}(-z/2)\Gamma\psi(z/2)\mid p,\Lambda\rangle_{z^+=0} 
\end{equation}
the parametrization of the new correlator gives us GTMDs \cite{MMS}, 
\begin{align}
\mathcal{W}^{\gamma^+}_{\Lambda,\Lambda'} &= \frac{1}{2M} \bar{u}(p',\Lambda') [F_{11} + \frac{i\sigma^{i+}k^i_T}{\bar{p}_+}F_{12} + \frac{i\sigma^{i+}\Delta^i_T}{\bar{p}_+}F_{13} + \frac{i\sigma^{ij}k^i_T\Delta^j_T}{M^2}F_{14} ]u(p,\Lambda) \\
&=\bigg(F_{11} +i\Lambda \frac{\vec{k_T}\times\vec{\Delta_T}}{M^2}F_{14} \bigg)\delta_{\Lambda\Lambda'} + \left(\frac{\Lambda \Delta^1 +i\Delta^2}{2M}(2F_{13} - F_{11}) + \frac{\Lambda k^1 +ik^2}{M}F_{12} \right)\delta_{\Lambda-\Lambda'}
\end{align}

\noindent If we look at the proton helicity non flip case ($\delta_{\Lambda\Lambda'}$) and, following (\ref{oameq}), take the difference of the proton helicity along $+$ and $-$ directions, the GTMD $F_{14}$ survives. It is interesting to note that at leading twist both the off forwardness and inclusion of transverse momentum of the partons is necessary to access OAM.

\noindent In \cite{Polyakov,Penttinen:2000dg} Polyakov et al showed that the twist three GPD $G_2$ connects to OAM (later also shown in \cite{Hatta}), 
\begin{equation}
W^{\gamma_\perp^i }(x,\zeta, t) = \bar{U}(P',\Lambda')\left[(H + E)\gamma_\perp^i + \frac{\Delta^i_\perp}{2M}G_1 + \gamma^i_\perp G_2 + \frac{\Delta_\perp^i \gamma^+}{P^+}G_3 + i\epsilon^{ij}_\perp \Delta^\perp_j \frac{\gamma^+\gamma^5}{P^+}G_4\right]U(P,\Lambda)
\end{equation}

\begin{equation}
-\int_{-1}^1dx x G_2(x,0,0)  = \frac{1}{2}\int_{-1}^1dx x(H_q(x,0,0) + E_q(x,0,0)) - \frac{1}{2}\int_{-1}^1dx \tilde{H}_q(x,0,0)\\
\end{equation}

\noindent The GPD $\tilde{E}_{2T}$ defined in \cite{MMS} is equivalent to $G_2$ and for uniformity, we will refer to this GPD as $\tilde{E}_{2T}$ in this text. To get more insight, we can look at the GTMD structure for $W^{\gamma^i}$ :
\begin{align}
\mathcal{W}^{\gamma^i}_{\Lambda\Lambda'}&= \frac{1}{P^+}\bigg(k_T^i F_{21} + \Delta_T^i F_{22} + i\Lambda \epsilon^{ij}_T(\frac{k_T^i}{M}F_{27} + \frac{\Delta_T^i}{M}F_{28}) \bigg)\delta_{\Lambda\Lambda'}
+ \frac{1}{P^+}\bigg(M(\Lambda\delta_{i1}+\delta_{i2})F_{23} \\ &+\frac{\Lambda k^1 +ik^2}{M}(k^i_TF_{24} + \Delta^i_TF_{25}) + \frac{\Lambda \Delta^1 +i\Delta^2}{2M}(2\Delta^i_TF_{26} - (k_T^i F_{21} + \Delta_T^i F_{22}))\bigg)\delta_{\Lambda-\Lambda'}
\end{align}

\noindent The GTMDs $F_{27}$ and $F_{28}$ survive when we subtract the proton $-$ and $+$ helicity. These are also exactly the GTMDs that contribute to $\tilde{E}_{2T}$.
\begin{equation}
2\int d^2k_T \frac{k_T.\Delta_T}{\Delta_T^2}F_{27} + F_{28} = \tilde{E}_{2T}
\end{equation}
\noindent Two GPDs H and E enter the parametrization at twist three in an analogous way as the PDF $g_1(x)$ enters the definition of higher twist PDF $g_T(x) = g_1(x) + g_2(x)$ \cite{Mulders,everybody}. Hence, the GPD $\tilde{E}_{2T}$ is the vector counterpart of $g_T(x)$ for the off forward case. The combination that contributes to $\tilde{E}_{2T}$ or $F_{27}$ and $F_{28}$ is 
\begin{equation}
W^{\gamma^1}_{++} -iW^{\gamma^2}_{++} -W^{\gamma^1}_{--} +iW^{\gamma^2}_{--}
\end{equation}

\noindent As we are looking now at the twist three scenario, the quark operators ($\gamma^i$ in this case) project out the ``bad'' quark field components which is a quark gluon combination \cite{JaffeSpin}.  Since  the gluon has spin one, the helicity of the quark is opposite to that of the bad component. We find that the twist three helicity amplitudes that correspond to $G_2$ are given by \cite{us2},
\begin{equation}
A_{+-^*,++} - A_{++,+-^*} - A_{--^*,-+} + A_{-+,--^*}
\end{equation}
\noindent Here, the starred helicities represent the bad component of the quark field. Due to the mixing of good and bad components, the interpretation of the helicity amplitudes as a probability distribution is more complicated at twist three. 


\section{Diquark Model Calculations}
\noindent The helicity amplitudes give us enormous physical insight and are very useful to perform model calculations. We use the parametrization of the spectator diquark model from \cite{GGL}. The initial proton dissociates into a quark and a recoiling fixed mass system with the quantum numbers of a diquark. The coupling at the 
the proton-quark-diquark vertex is given by,
\begin{figure}
\includegraphics[width=0.5\textwidth]{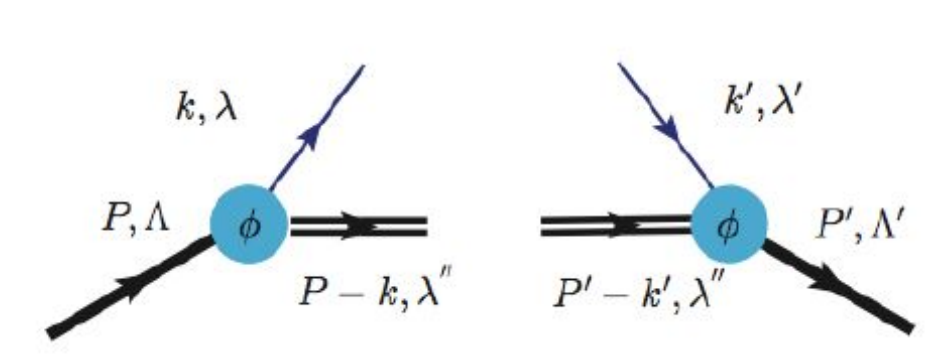}
\includegraphics[width=0.5\textwidth]{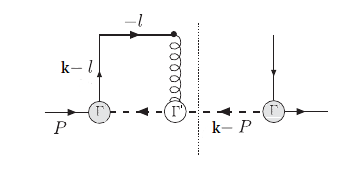}
\caption{Tree Level Spectator Diquark Model \cite{GGL}(\textit{left}) Interference between gluon exchange diagram and tree level diagram \cite{bac_con_rad}(\textit{right})}
\label{picture}
\end{figure}
\begin{equation}
\Gamma = g_s\frac{k^2 - m^2}{(k^2 -m_\Lambda^2)^2}
\end{equation}
\noindent $g_s$ being a constant. Although this is a scalar coupling, both scalar and vector diquark configurations can be achieved by varying the mass parameters. For the GTMDs, the quark - proton helicity amplitudes are defined as :

\begin{equation} 
\mathcal{A}_{\Lambda \lambda, \Lambda'\lambda'} = \phi_{\Lambda'\lambda'}^*(k',P')\phi_{\Lambda\lambda}(k,P)
\end{equation} 

\noindent where $\phi_{\Lambda\lambda}(k,P)$ and $\phi_{\Lambda'\lambda'}^*(k',P')$ are given by 

\begin{align}
\phi_{\Lambda\lambda}(k,P) &= \Gamma(k)\frac{\bar{u}(k,\lambda)U(P,\Lambda)}{k^2 - m^2} \\
\phi^*_{\Lambda'\lambda'}(k',P') &= \Gamma(k')\frac{\bar{U}(P',\Lambda')u(k',\lambda')}{k'^2 - m^2} 
\end{align}

\noindent For the case $\Delta^+=0$,

\begin{align}\label{phis}
\phi_{\Lambda\Lambda}(k,P) &= \frac{\Gamma(k)}{D\sqrt{x}}\bigg[(m+xM) + \frac{1}{2P^+}\bigg(\frac{-(m+xM)^2}{x} +M^2(1+x) +2mM - \frac{M_x^2 + k_\perp^2}{1-x}\bigg)\bigg] \\
\phi_{\Lambda-\Lambda}(k,P) &=  -\frac{\Gamma(k)}{D\sqrt{x}}(\Lambda k^1+ik^2)\bigg(1 +\frac{m+xM}{2xP^+}\bigg)\\
\phi^*_{\Lambda'\Lambda'}(k',P') &=\frac{\Gamma(k')}{D'\sqrt{x}}\bigg[(m+xM) + \frac{1}{2P^+}\bigg( 2mM+ M^2(1+x) - \frac{M_x^2 + k_T^2}{1-x}- \frac{(m+xM)^2}{x}\\ &+2k_T.\Delta_T  -\Delta_T^2(1-x) + 2i\Lambda'\Delta_T\times k_T \bigg)\bigg]     \\
\phi^*_{\Lambda'-\Lambda'}(k',P') &=  \frac{\Gamma(k')}{D'\sqrt{x}}(-\Lambda \tilde{k}^1+i\tilde{k}^2)(1 -\frac{m+xM}{2xP^+})
\end{align}

\noindent where, $\tilde{k} = k - (1-x)\Delta$, $D = k^2 - m^2$ and $D' = k'^2 - m^2$. In general, we can represent $\phi \rightarrow \phi^{Tw2} + \phi^{Tw3}$ i.e. as a sum of a twist two and a suppressed twist three term. At twist two we only use the leading order terms. The terms suppressed by $P^+$ represent the bad component of the quark field and enter at twist three. To calculate $F_{14}(x)$ we look at the amplitude combination in (\ref{oameq}) and using (\ref{phis}), arrive at the following, 
\begin{equation}
F_{14} = \mathcal{N}\frac{M^2}{x(k^2-m_\Lambda^2)^2((k-\Delta)^2-m_\Lambda^2)^2}.
\end{equation}
\noindent Next, we include final state interactions between the active quark and the spectator diquark \cite{burkardt_staple}. Similarly to the calculation of the Sivers function in forward kinematics \cite{bac_con_rad}, this is achieved by by considering the interference between the tree-level scattering amplitude and the single-gluon-exchange scattering amplitude in the eikonal approximation where the gluon momentum is almost all along the light cone (Fig.\ref{picture}, right),
\begin{equation}
F_{14} = \mathcal{N'}\frac{M^2(1-x)}{x(k^2-m_\Lambda^2)^2}\int \frac{d^2l_T}{(2\pi)^2}\frac{(1+\frac{l_T.k_T}{l_T^2})}{((l_T-k_T)^2-L_x^2(m^2) -2x l_T^2)^2}.
\end{equation}
\noindent Here, $\mathcal{N}$ and $\mathcal{N'}$ are normalization constants and $l$ represents the gluon momentum. $L_x^2(m^2) = xM_x^2 +(1-x)m^2 -x(1-x)M^2$ where, $M_x$ is the diquark mass.

\noindent The calculation of GPDs and PDFs involves helicity amplitudes where the partonic transverse momentum is integrated over,

\begin{equation}
A_{\Lambda \lambda, \Lambda'\lambda'} = \int d^2 k_T \mathcal{A}_{\Lambda \lambda, \Lambda'\lambda'}.
\end{equation} 

\begin{figure}
\centering
\includegraphics[width=0.5 \textwidth]{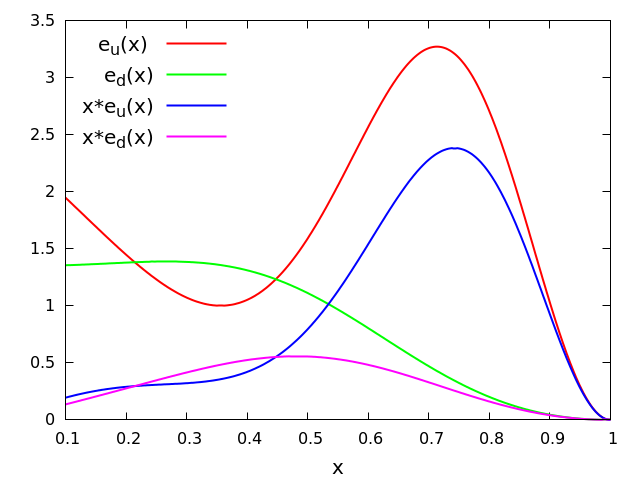}
\caption{$e(x)$ and $xe(x)$ for u and d quarks in diquark spectator model}
\label{fig:ex}
\end{figure}

\noindent For twist three calculations we need to consider interference of $\phi^{Tw2}$ and $\phi^{Tw3}$ terms,

\begin{align}
A_{\Lambda \lambda^*, \Lambda'\lambda'} &= \phi^{*Tw2}_{\Lambda'\lambda'}\phi^{Tw3}_{\Lambda\lambda} \\
A_{\Lambda \lambda, \Lambda'\lambda^{*'}} &= \phi^{Tw3}_{\Lambda'\lambda'} \phi^{*Tw2}_{\Lambda\lambda},
\end{align} 
where the starred notation on the {\it lhs} represents the bad components of the quark field. 

\noindent In Figure \ref{fig:ex} we show, as an example the twist three PDF, $e(x)$, that parametrizes the scalar component of the correlation function. $e(x)$ corresponds to the following helicity amplitude combination,
\begin{equation}
A_{++^*,++} + A_{++,++^*} + A_{+-^*,+-} + A_{+-,+-^*}+A_{-+^*,-+} + A_{-+,-+^*} + A_{--^*,--} + A_{--,--^*}.
\end{equation}

\noindent Results for the GPD $\tilde{E}_{2T}$ calculated using (\ref{oameq}) will be presented in a forthcoming publication \cite{us2}. In this case it is important to notice that the function's normalization can be fixed using the connection to the twist three reduced matrix element $d_2$ \cite{Polyakov,Jaffe_g2,Burkardt},
\begin{equation}
\int dx x^2\tilde{E}_{2T}^{Tw3} = -\frac{2}{3}d_2,
\end{equation}
\noindent where $\tilde{E}_{2T}^{Tw3}$ is the genuine twist three contribution to $\tilde{E}_{2T}$. Lorentz invariant relations can be derived between the $k_T^2$ moment of the twist two GTMDs and twist three GPDs\cite{everybody}. A key step is to look at the substructure of these functions with respect to functions that parametrize the non integrated correlator ($k^-$) aka Generalized Parton Correlation Functions (GPCFs). Our approach is very similar to that in \cite{Mulders,everybody}. The relation we find is the following,



\begin{equation}
\int d^2 k_T \frac{k_T^2}{M^2} F_{14}(y,k_T^2,0,0,0) = -\int_x^1 dy \tilde{E}_{2T}(y,0,0) + H(y,0,0) + E(y,0,0) \label{LIR}
\end{equation}

\noindent The moment in transverse momentum $k_T$ is important. In some sense, it mimics the quark gluon interactions that would be necessary to generate orbital angular momentum in a collinear picture.

\section{Conclusions}
\noindent The diquark model calculations of $F_{14}$ and $\tilde{E}_{2T}$ allow us to study both the Jaffe Manohar and Ji definitions of orbital angular momentum. Inclusion of final state interactions in the picture is relevant for OAM and for other non-perturbative effects such as single spin asymmetries in the proton.  We have derived a relation between the GTMD (via $F_{14}$) and GPD ($\tilde{E}_{2T}$) definitions of OAM. 
It is essential that these objects be measured experimentally and this is something we are currently working on. Much interesting physics remains to be explored in the future including the evolution of these functions, and the role of the genuine twist three contribution.

\end{document}